# THERE IS AN ANSWER

## By Rodney A. Brooks

(author of "Fields of Color: The theory that escaped Einstein")

**Abstract: Recently there has been an explosion of books and articles complaining about the weirdness of Quantum Mechanics and crying out for a solution. Three problems in particular have been singled out: the double-slit experiment, the measurement problem, and entanglement. One of these (entanglement) was the subject of an episode of the BBC TV show NOVA. In this article it is shown that Quantum Field Theory, as formulated by Julian Schwinger, provides simple solutions for all three problems, and others as well.**

## INTRODUCTION

Physicists have struggled with Quantum Mechanics and its paradoxes for 90 years:

> **I think I can safely say that nobody understands quantum mechanics. — R. Feynman[1]**

Lay people have also struggled:

> **In recent years I have been… writhing in an exasperating quandary over quantum mechanics, which, to my mind, remains impossible even to define, let alone comprehend. – J. Heller[2]**

One physicist, Paul Ehrenfest, was driven to suicide because of it:

> **In recent years it has become ever more difficult for me to follow the developments in physics with understanding. After trying, ever more enervated and torn, I have finally given up in desperation. This made me completely weary of life... I have no other practical possibility than suicide… — P. Ehrenfest[3]**

Most physicists handle this problem by giving up hope of understanding QM. "Shut up and calculate" has become the rule.

> **I am a positivist who believes that physical theories are just mathematical models we construct, and that it is meaningless to ask if they correspond to reality, just whether they predict observations. – S. Hawking[4]**

Today, however, this is beginning to change. Four books were published in 2018, all making the point that Quantum Mechanics doesn't make sense and that an answer is needed.[5,6,7,8] One of them (*Beyond Weird*) won the Physics World Book of the Year award. Three problems in particular are singled out: The double-slit experiment, the measurement problem, and entanglement. One of these problems (entanglement) has also been featured on public television,[9] and even satirized in the comics pages.[10]

# THE PROBLEMS

Double-slit experiment.  In 1801 Thomas Young showed that a beam of light forms an interference pattern after passing through two slits, thereby proving that light is made of waves.  Later, when the quantum nature of light was discovered, the same experiment was repeated with individual photons and the same pattern was seen after a large number of individual photons were detected, showing that each photon interferes with itself.  The double-slit experiment has also been performed with electrons, an experiment that Richard Feynman said couldn't be done.

> **"Now we imagine a similar experiment with electrons… We should say right away that you should not try to set up this experiment… The trouble is that the apparatus would have to be made on an impossibly small scale to show the effects we are interested in."**
> **– R. Feynman** [11]

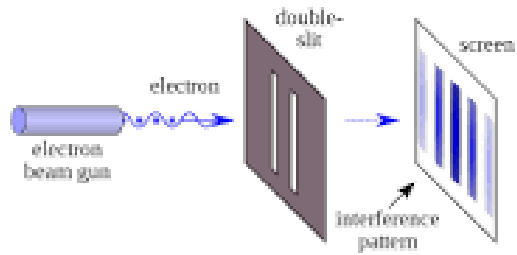

Fig. 1.  The double-slit experiment performed with electrons.

The "impossible" experiment was carried out in 1961 and was repeated in 1974 with an electron source so weak that the detection of single electrons could be registered (Fig. 1).  It was voted "the most beautiful experiment" by readers of *Physics World*.[12]

The double-slit experiment is the quintessential example of wave-particle duality.

> **[Wave-particle duality] is a paradox because particles are, by definition, localized entities that follow definite trajectories while waves are not confined to any particular path or region of space.  How could the same thing be both confined and not confined, both a particle and a wave? – T. Norsen ("Intelligent design in the classroom",** *Forum on Physics and Society 35*, **3, 2006)**

This paradox has led to the joke: "light is waves on Mondays, Wednesdays and Fridays; it is particles on Tuesdays, Thursdays and Saturdays; and on Sundays we think about it."  What is needed is a theory that shows how a photon or electron can behave like a wave and yet be detected like a particle on every day of the week.

Measurement problem.  The measurement problem has been called "the most controversial problem in physics today".[13]  It began in 1935 when Albert Einstein wrote to Erwin Schrödinger to express frustration with the idea that a quantum system is in a superposition of states until an observation is made. To show how ridiculous this is, Einstein suggested a hypothetical experiment in which a pile of gunpowder is connected to an unstable particle so that when the particle decays the gunpowder will combust.  "A *sort of blend of not-yet and already-exploded systems*", said Einstein, "*[cannot be] a real state of affairs, for in reality there is just no intermediate between exploded and not exploded.*" [14]  Schrödinger wrote back saying "*In a lengthy essay that I have just* *written, I give an example that is very similar to your exploding powder keg.*"[15] Schrödinger's example was a cat enclosed in a chamber with a radioactive sample and a Geiger counter that is connected through a relay to a flask of poison gas. After a certain time has passed, he wrote, "*the [wave-function] of the entire system would express this by having in it the living and dead cat… mixed or smeared out in equal parts*" (Fig. 2.). Schrödinger called such a scenario "ridiculous".

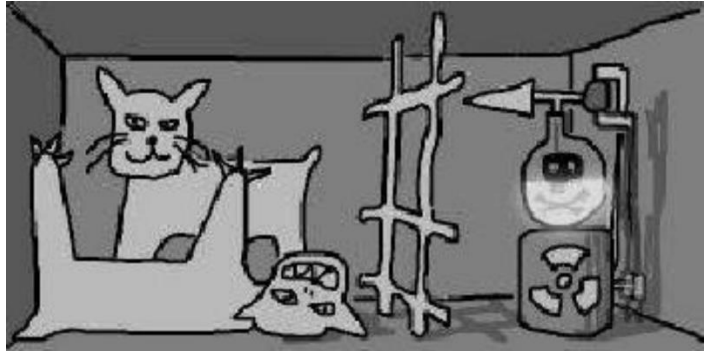

Fig. 2. Schrödinger's experiment with the cat both alive and dead.

It is not just the superposition of a cat or a bomb that Schrödinger and Einstein were satirizing. These were clearly a *reductio ad absurdum* to show that the very idea of superpositions, whether microscopic or macroscopic, makes no sense. If a cat that is half dead and half alive is not believable, then neither is a relay that is both tripped and untripped, or a Geiger counter that is both discharged and undischarged, or even an atom that is both ionized and unionized. What is needed is a theory that offers a believable picture of reality at every moment.

Entanglement. If two photons are created together so that their properties are correlated, a measurement made on one photon affects the same property in the other photon. What's more, the effect is instantaneous, no matter how far apart are the photons. Such photons are said to be *entangled*. Entanglement has been demonstrated over distances as great as 1200 kilometers,[16] yet, despite the experimental proof, physicists have a hard time accepting that it happens. Einstein famously called it "spooky action at a distance" and it was the subject of a NOVA TV episode entitled "Einstein's Quantum Riddle." As Steven Weinberg wrote

> **Strange as it is, the entanglement entailed by quantum mechanics is actually observed experimentally. But how can something so non-local represent reality? – S. Weinberg[17]**

What is needed is a theory that makes non-locality plausible.

## QUANTUM FIELD THEORY

In 1949, after completing his Nobel-award winning work on renormalization,[18] Julian Schwinger felt that a new approach was needed:

> **"The pressure to account for those [experimental] results had produced a certain theoretical structure that was perfectly adequate for the original task, but demanded simplification and generalization; a new vision was required… Like the silicon chip of more recent years, the Feynman diagram was bringing computation to the masses… But eventually one has to put it all together again, and then the piecemeal approach loses some of its attraction… Quantum field theory must deal with Bose-Einstein fields and Fermi-Dirac fields on a fully equivalent footing… There was my challenge." – J. Schwinger** [19]

Schwinger's challenge led to a series of six papers titled "A theory of quantized fields".[20] In this theory there are no particles and there are no superpositions of states; there are only fields and field quanta. However Schwinger's formulation of QFT is virtually unknown today, while it is Richard Feynman's version, based on particles and virtual particles, that is generally accepted. Therefore, we will start with a brief review.

Fields. A field is a property of space. This concept was introduced by Michael Faraday in 1845 to explain electric and magnetic forces, and in 1864 James Maxwell developed equations to describe the evolution of these fields. However it took many years before physicists were able to drop the idea that electromagnetic forces were carried by an ether and to accept them as properties of space. As Paul Drude wrote in 1900:

> **"The conception of an ether absolutely at rest is the most simple and most natural – at least if the ether is conceived to be not a substance but merely space endowed with certain physical properties."** [21]

Discreteness. In 1900 Max Planck showed that the EM field is made of discrete units, and in 1922 the Stern-Gerlach experiment showed that a physical quantity (angular momentum) can have only discrete values. Mathematically, a quantity that can have only $n$ discrete values is described using a complex vector space of $n$ dimensions, called Hilbert space. The discrete values of the physical quantity are called *eigenvalues* and for each eigenvalue there is a corresponding *eigenvector* in Hilbert space. These eigenvectors form an orthonormal set and the value of the physical property is represented by a vector in this space - but not necessarily an eigenvector: it can be a superposition of eigenvectors. For this reason one does not talk about *the* value of the property but about its *expectation value*. Finally, there are *operators* in Hilbert space that act on the state vectors and are used to describe the dynamics of the physical property.

Field quantization. In QFT the concept of discreteness is extended to field strength.[22] That is, the continuum of values is regarded as the limiting case of many discrete values and the corresponding Hilbert space is expanded to an infinite number of dimensions. However the physical fields that exist in real three-dimensional space should not be confused with their mathematical representation in Hilbert space, and Hilbert operators should not be confused with the real fields.

Quanta. The use of Hilbert algebra leads to the existence of units of field called *quanta* (a term introduced by Planck). Each quantum is spread out in space but acts as a unit; it is a separate entity that lives a life and dies a death of its own. Quanta can be either free or bound. Examples of free quanta are a photon emitted by a lamp or an electron emitted from a cathode. Examples of bound quanta are the electrons in an atom or the nucleons in a nucleus. If bound quanta acquire enough energy they can become free, but as long as they are bound together they must remain together. In addition to quanta, there are also *self-fields*, such as the EM field surrounding an electron or the strong force field surrounding a nucleon. Finally, there is a *vacuum field* that is present even if there are no quanta.

Field equations. These fields evolve and interact as specified by equations for the corresponding operators. As Frank Wilczek said,

> **"The move from a particle description to a field description will be especially fruitful if the fields obey simple equations, so that we can calculate the future values of fields from the values they have now… Evidently, Nature has taken the opportunity to keep things relatively simple by using fields**." [23]

Quantum collapse. The field equations do not tell the whole story; in fact, they don't tell the most important part of the story. They don't describe transfer of energy, and without energy transfer nothing of significance can happen, including measurement. When a quantum transfers its energy to an atom it must collapse into that atom; it cannot continue to exist without energy. Even if a quantum transfers part of its energy, the entire quantum must collapse and be re-emitted. As Art Hobson wrote, referring to tracks in a cloud chamber:

> **"The tracks are made by successive individual interactions between a matter field and gas or water molecules. The matter quantum collapses… each time it interacts with a molecule, while spreading out as a matter field between impacts."** [24]

Other examples of quantum collapse are a photon depositing its energy into a photoreceptor in the eye or a radiated quantum transferring energy to an atom in a Geiger counter. Collapse of a different kind occurs when an internal property of a quantum, such as spin, is changed. The spin cannot have one value in one part of the quantum and a different value in another part, and the change must be instantaneous throughout the quantum. Unlike the evolution described by the field equations, quantum collapse is irreversible.

Probability remains. Since there is no theory to describe quantum collapse, we can't predict when or where it will happen. All we know is that the probability of collapse is related to the field strength at that point. This is not the first time in physics history that something was known to happen without a theory to explain it. Eventually we may find a theory that describes quantum collapse, but in the meantime we must accept it as a necessary corollary of quantized fields.

Non-locality. If a quantum collapses, it disappears from all space instantly, no matter how spread out it is. This *non-locality* is a hard pill for many to swallow, but it is a necessary corollary of quantized fields. If its energy is gone, a quantum cannot continue to exist and be available to transfer energy to another atom. Besides, non-locality is an experimental fact,[25] and there is nothing logically inconsistent about it.

Einstein famously refused to believe that a field spread out over millions of light-years could instantaneously disappear. Dirk Bouwmeester's answer to this was "In for a penny, in for a pound."[26] In other words, it doesn't matter if the distance is a nanometer or a light-year – if you can accept one, you should be able to accept the other.

Non-locality also violates the principle of Relativity, because what is simultaneous in one frame of reference is not simultaneous in another. However this does not lead to any inconsistency. Removing a field before it has had a chance to do anything is not the same as changing something that has already happened, and changing the spin (or other property) of a field before it has had a chance to do something is not the same as changing something that has already happened.

Source theory. Quantum collapse was not included in Schwinger's original formulation of QFT, but it became an important part of the *source theory* that he developed later. (In the following passage the word "particle" is used to mean field quantum.)

> **"We have spoken of particle creation, but equally important is particle detection… In a general sense the particle is annihilated by the process of detecting it. The… processes used to detect a particle can be idealized as sinks wherein the particle's properties are handed on… but sink and source are clearly different aspects of the same idealization, and we unite them under the general heading of sources."** [27]

That source theory was not accepted caused Schwinger great disappointment:

> **"Developments…. in which the new viewpoint was successfully applied convinced me, if no one else, of the great advantages that its use bestowed. The lack of appreciation of these facts by others was depressing."** [28]

QFT vs. QM. QFT and Quantum Mechanics are very different theories, but the difference is not so much in the equations. After all, the Schrödinger equation of QM is the non-relativistic limit of the QFT equation for matter fields. The difference is in their interpretation. The Schrödinger equation gives the *probability* that a particle is at a given point while the QFT equations give the *field strength* at that point, as represented by a vector in Hilbert space. Wave-function collapse in QM is a collapse of probabilities, but in QFT quantum collapse is a physical event.

We will now show how this theory provides answers to the problems described above, and to other problems of physics as well.

THE ANSWERS

The double-slit experiment. In QFT, photons and electrons are field quanta. The quanta spread out in space, as only a field can do, and therefore pass through both slits, but when they reach the detector, they collapse into only one atom. Since the probability of collapse is related to the field strength, the classic Young interference pattern is seen at the detector. There is no wave-particle duality paradox.

> **…these two distinct classical concepts [particles and fields] are merged and become transcended in something that has no classical counterpart – the quantized field that is a new conception of its own, a unity that replaces the classical duality. – J. Schwinger**[29]

The double-slit experiment has even been performed with molecules, and the QFT picture is similar. The components of the molecule (electrons, protons and neutrons) are all quanta of fields that are bound together. If one collapses, they must all collapse together. One quantum cannot collapse and leave the other bound quanta to go free.

The measurement problem. Schrödinger's cat experiment can be divided into two stages. During the initial stage the radiated quantum spreads throughout space and interacts with all quanta that it encounters, in accordance with the field equations. These interactions are deterministic and reversible. This phase ends when the quantum collapses and transfers energy to an atom. If the absorbing atom happens to be in a Geiger counter, it initiates a Townsend discharge that trips the relay that releases the gas that kills the cat. Until then the cat is alive; after that the cat is dead.

While QFT does not predict the moment or place of collapse, it provides a realistic picture of what is happening at every moment — in fact, the same picture that people had before QM came along.

Entanglement. The entanglement experiment is another instance — a more elaborate instance — of quantum collapse. If one can accept that a single quantum, spread over miles (or light-years) of space, can instantaneously collapse, it is not much of a stretch to accept that two entangled quanta can do the same. In fact, if one quantum collapses, conservation of momentum *requires* that the other one also collapse. However this secondary collapse is not the same as the primary collapse: It is not determined by field intensity, but is a consequence of the primary collapse.

Not only is the entanglement experiment explained by QFT, it also provides proof that non-locality is real – that it cannot be denied. Whether the quanta are seen as fields or particles, when one is detected, the state of the other quantum is known instantaneously, no matter how far apart they are.

Besides the above problems, QFT resolves many other problems and mysteries of QM. In addition, it provides simple explanations for the paradoxes of Relativity. We will now look at some examples.

The delayed-choice experiment. The delayed-choice experiment, proposed by John Wheeler in 1983,[30] is based on the Mach-Zehnder interferometer (Fig. 3). In the upper picture, a photon enters at the lower left and passes through a beamsplitter that sends each beam along a different path. Each path leads to a detector (not shown), and according to the conventional interpretation, the choice of detector indicates which path the photon has taken.

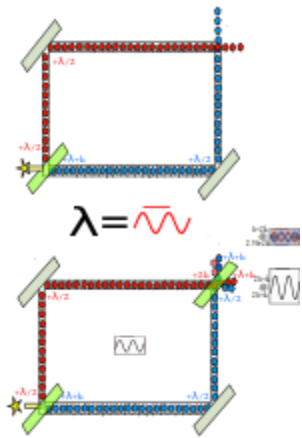

Fig. 3. Wheeler's delayed choice experiment (from Wikipedia).

In the "closed" configuration (lower picture) there is a second beamsplitter (upper right) that combines the two beams and sends the combination to each detector. It also changes the phase of one of the reflected beams by 180º, so there is destructive interference at one detector and constructive interference at the other  However the phase difference between the two paths can be varied, so the interference effect between detectors will depend on the relative phases, showing that the photon has followed both paths. In other words, in the open configuration the photon acts like a particle and in the closed configuration it acts like a wave. Wheeler wondered what would happen if the second beamsplitter is inserted after the "choice" (particle or wave) is made.

The experiment was carried out in 2006[31] and the result was as Wheeler had predicted. How can the photon "decide" whether to travel both routes or only one route after it has passed the dividing point?

The QFT explanation couldn't be simpler. The photon is a field. It *always* travels along both routes and it *always* impinges on both detectors. It then collapses into an atom in one of the detectors with a probability that depends on the field strength at that point. This explains why, in the open configuration, half the time detector A is triggered and the other half detector B is triggered, with the choice being random, but if the recombining device is inserted, there is interference between the two parts that affects the choice of detector.

The Uncertainty Principle. The *uncertainty principle*, formulated by Werner Heisenberg in 1927, says that the exact position of a particle cannot be determined, but that the uncertainty in position is related to the uncertainty in momentum. Now the spatial spread of any field is related to the spread in wavelength by Fourier's Theorem, and in QFT the wavelength is related to momentum. Thus Heisenberg's Uncertainty Principle for particles is nothing more than Fourier's theorem for fields.

The Exclusion Principle. The *exclusion principle* was introduced by Wolfgang Pauli in 1925. It states that two or more electrons (or other fermions) cannot occupy the same quantum state at the same time. The reason for this didn't come until the spin-statistics theorem was derived from early versions of QFT.

> **In my original paper I stressed the circumstance that I was unable to give a logical reason for the exclusion principle or to deduce it from general assumptions… If we search for a theoretical explanation of this law, we must pass to the discussion of relativistic wave mechanics. – W. Pauli (Nobel lecture, 1945)**

The sub-atomic zoo. During the late twentieth century the number of hadrons multiplied rapidly, resulting in what is called the *sub-atomic zoo*. In a recent tour de force of computer technology, it was found that the equations for the fundamental quark and gluon fields successfully predict the mass and spin of these hadrons.[32] To perform the calculation, the basic parameters of the quark and gluon fields were first determined from three known hadron masses, after which stable and quasi-stable excitations in the fields were calculated. Fourteen such excitations were found with mass and spin in close agreement with the fourteen known hadrons — ranging from protons and neutrons to the exotic charmonium. Equally important is that there were no excitations corresponding to the quarks and gluons themselves, thereby providing a theoretical basis for the *Principle of Confinement*. Frank Wilczek called this "*one of the greatest scientific achievements of all time.*" [33]

Relativity paradoxes. QFT also offers a "bottom-up" explanation for the counter-intuitive results of Special Relativity, such as time dilation, length contraction, etc. It shows how these paradoxical results are a natural consequences of the way fields behave.[34]

CONCLUSION

Quantum Field Theory is a self-consistent theory that not only agrees with, but explains, many results of quantum experiments. The primary objection to QFT is the non-locality of quantum collapse. However non-locality is an experimental fact and presents no internal contradictions. The real mystery is why this simple and elegant theory has been ignored for so long by the physics community.

**REFERENCES**

1. Feynman, R, *The Character of Physic al Law*, The MIT Press, Cambridge, MA. (1965)
2. Heller, J, *Now and Then*, Alfred A. Knopf, New York, NY. (1998). p. 194
3. https://medium.com/starts-with-a-bang/the-tragic-fate-of-physicist-paul-ehrenfest-93c946b05d0c
4. Penrose, R, et al, *The Large, the Small and the Human Mind*, Cambridge University Press, p. 169
5. Ball, P, *Beyond Weird: Why everything you thought you knew about quantum physics is different*, University of Chicago Press, Chicago, IL (2008)
6. Ananthaswamy, A, *Through Two Doors at Once: The elegant experiment that captures the enigma of our quantum reality*, Dutton, an imprint of Penguin Random House LLC, New York, NY. (2018)
7. Becker, A, *What Is Real? The unfinished quest for the meaning of quantum physics,* Basic Books, New York, NY. (2018)
8. Weinberg, S, "The problem with Quantum Mechanics", Chap. 4 of *Third Thoughts*, Harvard Univ. Press, Cambridge, MA. (2018)
9. "Einstein's Quantum Riddle", NOVA, January 9, 2016
10. Bell, D, "Candorville" comic strip, Sept. 4, 2016
11. Feynman R, Leighton R, and Sands, M, *The Feynman Lectures on Physics (vol 3)*, Basic Books, New York, NY (2015)
12. Crease, R, "The Most Beautiful Experiment," *Physics World*, September, 2002.
13. http://www.informationphilosopher.com/
14. Isaacson, W, *Einstein: His Life and Universe,* Simon & Schuster, New York, NY. (2007) p. 456
15. Schrödinger E, "The Present Status of Quantum Mechanics", *Die Naturwissenschaften*. **23,** p. 807-849 (1935)
16. Gao, WB, Lu, C-Y, Yao X-C, et al. *Experimental demonstration of a hype-entangled ten-qubit cat state. Nature Physics,* 6**:** p. 331 (2010)
17. Weinberg, S, op.cit. p. 137
18. Renormalization is a way to handle the infinities produced by interaction between a charged particle and its self-field, so that calculations can be made.
19. Schwinger J, "Two shakers of physics." In: Brown LM, Hoddeson L, editors. *The Birth of Particle Physics*. Cambridge, UK: Cambridge University Press; 1983. p. 343-344.
20. Schwinger J, The Theory of Quantized Fields I. Phys Rev 1951; **82**: 914.
    Schwinger J, The Theory of Quantized Fields II. Phys Rev 1953; **91**: 713.
    Schwinger J, The Theory of Quantized Fields III. Phys Rev 1953; **91**: 728.
    Schwinger J, The Theory of Quantized Fields IV. Phys Rev 1953; **92**: 1283.

Schwinger J, The Theory of Quantized Fields V. Phys Rev 1954; **92**: 615.
Schwinger J, The Theory of Quantized Fields VI. Phys Rev 1954; **93**: 1362

21. Drude P., quoted in *'Subtle is the Lord…': The science and the life of Albert Einstein* by Abraham Pais, Clarendon Press, Oxford (UK): Oxford Univ Press; 1982. p. 121

22. Schwinger J, "The Algebra of Microscopic Measurement." *Proc Natl Acad Sci USA* 1959; 45:1542

23. F. Wilczek, *The Lightness of Being: Mass, ether, and the unification of forces*. New York (NY): Basic Books; 2008, p. 86

24. Hobson A, *Physics: Concepts and Connections* (4th edit.). Pearson Prentice Hall, Upper Saddle River, NJ (2007) p. 300

25. Fuwa M, Takeda S, Schrödinger Zwierz M, Wiseman HW, Furusawa A. "Experimental proof of nonlocal wavefunction collapse for a single particle using homodyne measurements." *Nat Commun* 2015 Mar 24; 6: #6665

26. D. Bouwmeester, private conversation

27. Schwinger, J, *Particles, Sources, and Fields* vol 1. Reading (MA): Perseus Books; 1989. p. 38

28. Ibid. p. xi

29. Schwinger, J, *Quantum Mechanics: Symbolism of Atomic Measurements*, ed. Englert BG, Springer, Berlin. (2009). p. 4

30. Wheeler JA. "Law without law." In: Wheeler JA, Zurek WH, editors. *Quantum Theory and Measurement*. Princeton (NJ): Princeton Univ Press; 1983. p.182-213

31. Jacques V, Wu E, Grosshans F, et al. "Experimental realization of Wheeler's delayed-choice Gedanken Experiment." *Science* 2007; **315**: p. 966-968

32. Aoki S, Boyd G, Burkhalter R et al. "Quenched Light Hadron Spectrum". *Phys Rev Lett* 2000; **84**(2): p 238-41.

33. Wilczek F. *The Lightness of Being: Mass, ether, and the unification of forces*. New York (NY): Basic Books; 2008, p. 127

34. R.A. Brooks. *Fields of Color: The theory that escaped Einstein*. 3rd ed., Epsilon Publishers, Sedona (AZ), 2016